\documentclass[aps,prl,reprint,twocolumn,citeautoscript,superscriptaddress,showkeys,showpacs]{revtex4-1}   %
\usepackage{xr}
\usepackage[fleqn,tbtags]{amsmath}
\usepackage{amsfonts, amssymb,amsxtra,textcomp}
\usepackage[]{graphicx}
\usepackage{grffile}
\usepackage{bbm}
\usepackage{bm}
\usepackage{color}
\usepackage{graphicx}
\usepackage[colorlinks=true,citecolor=blue,linkcolor=blue]{hyperref}

\allowdisplaybreaks
\usepackage{pdfpages}

\newcommand{\av}[1]{\langle #1 \rangle}

\newcommand{\bfss}{{\boldsymbol{S}}}

\newcommand{\bfm}{{\boldsymbol{m}}}

\newcommand{\bft}{{\boldsymbol{t}}}

\def\bmx{\begin{pmatrix}}
\def\emx{\end{pmatrix}}

\renewcommand{\approx}{\simeq}

\begin{document}
\title{Emergent Power-Law Phase in the 2D Heisenberg Windmill Antiferromagnet: \\ A Computational Experiment}
\author{Bhilahari Jeevanesan}
\affiliation{Institute for Theory of Condensed Matter, Karlsruhe Institute of Technology (KIT), 76131 Karlsruhe, Germany}
\author{Premala Chandra}
\affiliation{Center for Materials Theory, Rutgers University, Piscataway, New Jersey 08854, USA}
\author{Piers Coleman}
\affiliation{Center for Materials Theory, Rutgers University, Piscataway, New Jersey 08854, USA}
\affiliation{Hubbard Theory Consortium and Department of Physics, Royal Holloway, University of London, Egham, Surrey TW20 0EX, UK}
\author{Peter P. Orth}
\affiliation{Institute for Theory of Condensed Matter, Karlsruhe Institute of Technology (KIT), 76131 Karlsruhe, Germany}
\affiliation{School of Physics and Astronomy, University of Minnesota, Minneapolis, Minnesota 55455, USA}
\date{\today}
\pacs{75.10.-b, 75.10.Jm}

\begin{abstract}
In an extensive computational experiment, we test Polyakov's conjecture that under certain circumstances
an isotropic Heisenberg model can develop algebraic spin 
correlations. We demonstrate the emergence of a multi-spin U($1$) order parameter in a Heisenberg antiferromagnet on interpenetrating honeycomb and triangular lattices. The correlations of this relative phase angle are observed to decay algebraically at intermediate temperatures 
in an extended critical phase. Using finite-size scaling, we show that both phase transitions are of the Berezinskii-Kosterlitz-Thouless 
type and at lower temperatures, we find long-range $\mathbb{Z}_6$ order. 
\end{abstract}
\maketitle

In statistical mechanics it is assumed~\cite{Polyakov_GaugeFieldsAndStrings_Book,zinn-justin_QFT} that 2D Heisenberg magnets cannot develop algebraic order at finite temperatures since interaction of the Goldstone modes causes the spin-wave stiffness to
renormalize to zero. However, in his pioneering work on this subject~\cite{polyakov_nlsm_75}, Polyakov speculated that a 2D
Heisenberg magnet might develop algebraic order if the
system were to develop a ``vacuum degeneracy''; he further 
suggested that this possibility might be explored experimentally. Recently
Orth, Chandra, Coleman and Schmalian (OCCS) have proposed that frustration
can provide a mechanism to realize Polyakov's conjecture; here fluctuations 
induce an emergent XY order parameter that decouples from the
rotational degrees of freedom~\cite{PhysRevLett.109.237205,PhysRevB.89.094417}.  However these arguments were based on a long-wavelength renormalization
group analysis, leaving open the possibility
that short-wavelength fluctuations could preempt the scenario
via unanticipated transitions into different
phases~\cite{Cardy80,PhysRevLett.106.207202,Chern12}. In this Letter,
we report a computational experiment that detects the
development of an emergent XY order parameter in a
2D Heisenberg spin model with power-law correlations, confirming the
OCCS mechanism and its realization of the Polyakov conjecture.

\begin{figure}[b!]
\includegraphics[width=\linewidth]{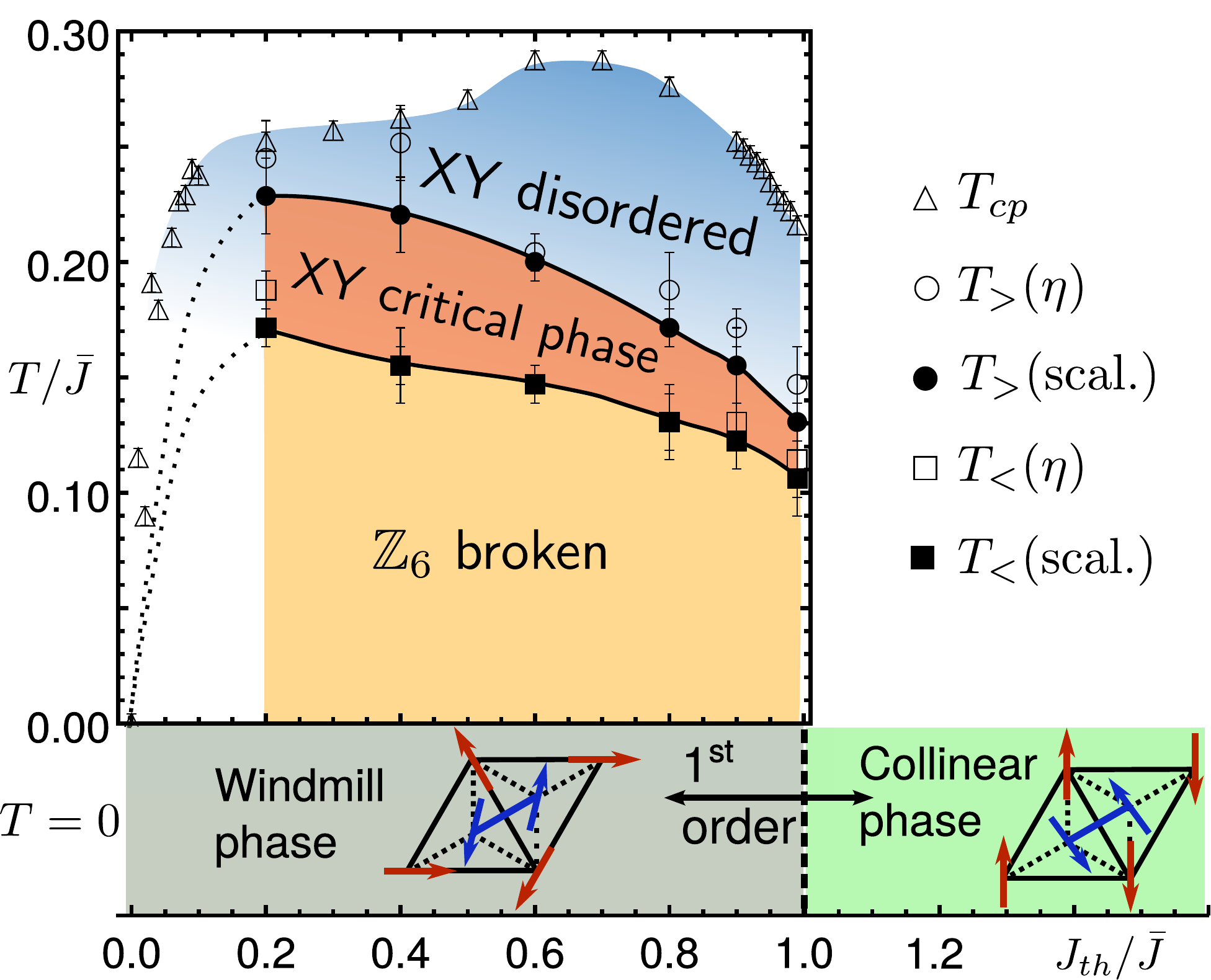} 
\caption{(color online). Finite temperature phase diagram of classical windmill Heisenberg antiferromagnet as a function of inter-sublattice coupling $J_{th}/\bar{J}$, $\bar{J} = \sqrt{J_{tt} J_{hh}}$. Below a coplanar crossover temperature $T_{cp}$, emergent XY spins appear and undergo two BKT phase transitions: at $T_>$ from a disordered to a critical phase with algebraic order and then at $T_<$ into a $\mathbb{Z}_6$ symmetry broken phase with discrete long-range order. At zero temperature the system undergoes a first order transition at $J_{th} = \bar{J}$ from a $120^\circ$/N\'eel ordered windmill phase to a collinear phase.  } 
\label{fig:1}
\end{figure}

The OCCS mechanism relies on the formation of a multi-spin U$(1)$
order parameter describing the {\sl relative} orientation of 
the magnetization between a honeycomb and a triangular lattice. 
The development of discrete multi-spin order is well known in systems with
competing interactions: an example is the fluctuation-induced
$\mathbb{Z}_{2}$ order in the $J_{1}-J_{2}$ Heisenberg
model~\cite{PhysRevLett.64.88}. This mechanism is thought to be
responsible for the high temperature nematic phase observed in the
iron-pnictides~\cite{PhysRevLett.105.157003,PhysRevB.85.024534,PhysRevLett.91.177202, PhysRevLett.92.157202}.  
In the OCCS mechanism, the emergent $U (1)$ order parameter is subject
to a $\mathbb{Z}_6$ order-by-disorder potential at short distances. 
At intermediate temperatures this potential is irrelevant (in the renormalization group sense) and scales to zero at long distances, leading to emergent
power-law correlations. Remarkably, the stiffness of the emergent
U$(1)$ order parameter remains finite in the infinite system, despite
the short-range correlations of the underlying Heisenberg spins. In
this XY manifold the binding of logarithmically interacting defect
vortices leads to multi-step ordering via two consecutive transitions
in the Berezinskii-Kosterlitz-Thouless (BKT) universality
class~\cite{PhysRevLett.109.237205,PhysRevB.89.094417,PhysRevB.16.1217}. 

The Hamiltonian studied by OCCS is the ``Windmill Heisenberg antiferromagnet'',
given by
$H = H_{tt} + H_{AB} + H_{tA} + H_{tB}$ with
\begin{align}
  \label{hamiltonian}
  H_{ab} &= J_{a b} \sum_{j = 1}^{N} \sum_{\{\delta_{ab}\}} \bfss^a_j \cdot \bfss^b_{j +\delta_{ab}} \,,
\end{align}
where $\bfss^a_j$ denote classical Heisenberg spins at Bravais lattice site $j$ and basis site $a\in \{t, A, B\}$. The windmill lattice can be described as interpenetrating and coupled triangular $(t)$ and honeycomb ($A,B$) lattices. The indices $\delta_{ab}$ relate nearest-neighbors of sublattices $a, b$, counting each bond once. The antiferromagnetic exchange couplings are $J_{tt}$, $J_{th} \equiv J_{tA} = J_{tB}$ and $J_{hh} \equiv J_{AB}$, and we introduce $\bar{J} = \sqrt{J_{tt} J_{hh}}$.

We employ large-scale parallel tempering classical Monte-Carlo simulations to obtain the finite temperature phase diagram shown in Fig.~\ref{fig:1}.
As the emergent order parameter is a multi-spin object, we had to design a specific non-local Monte-Carlo updating sequence consisting of three sub-routines: (i) a heat bath step~\cite{0022-3719-19-14-020} in which a randomly chosen spin is aligned within the local exchange field of its neighbors according to a Boltzmann weight; (ii) a standard parallel tempering move~\cite{0295-5075-19-6-002,doi:10.1143/JPSJ.65.1604} for which we run parallel simulations at $40$ temperature points and switch neighboring configurations according to the Metropolis rule; finally step (iii) is specifically tailored to our system where the emergent spins, defined below, exhibit a minute $\mathbb{Z}_6$ order-by-disorder potential. We select a (global) rotation axis perpendicular to the average plane of the triangular spins, which exhibit (local) $120^\circ$ order, and rotate all honeycomb spins around this axis by a randomly chosen angle and accept according to the Metropolis rule. This Monte Carlo algorithm was applied at least for $9 \times 10^5$ Monte-Carlo steps of which the first half is discarded to account for thermalization.

\begin{figure}[t!]
\includegraphics[width=\linewidth]{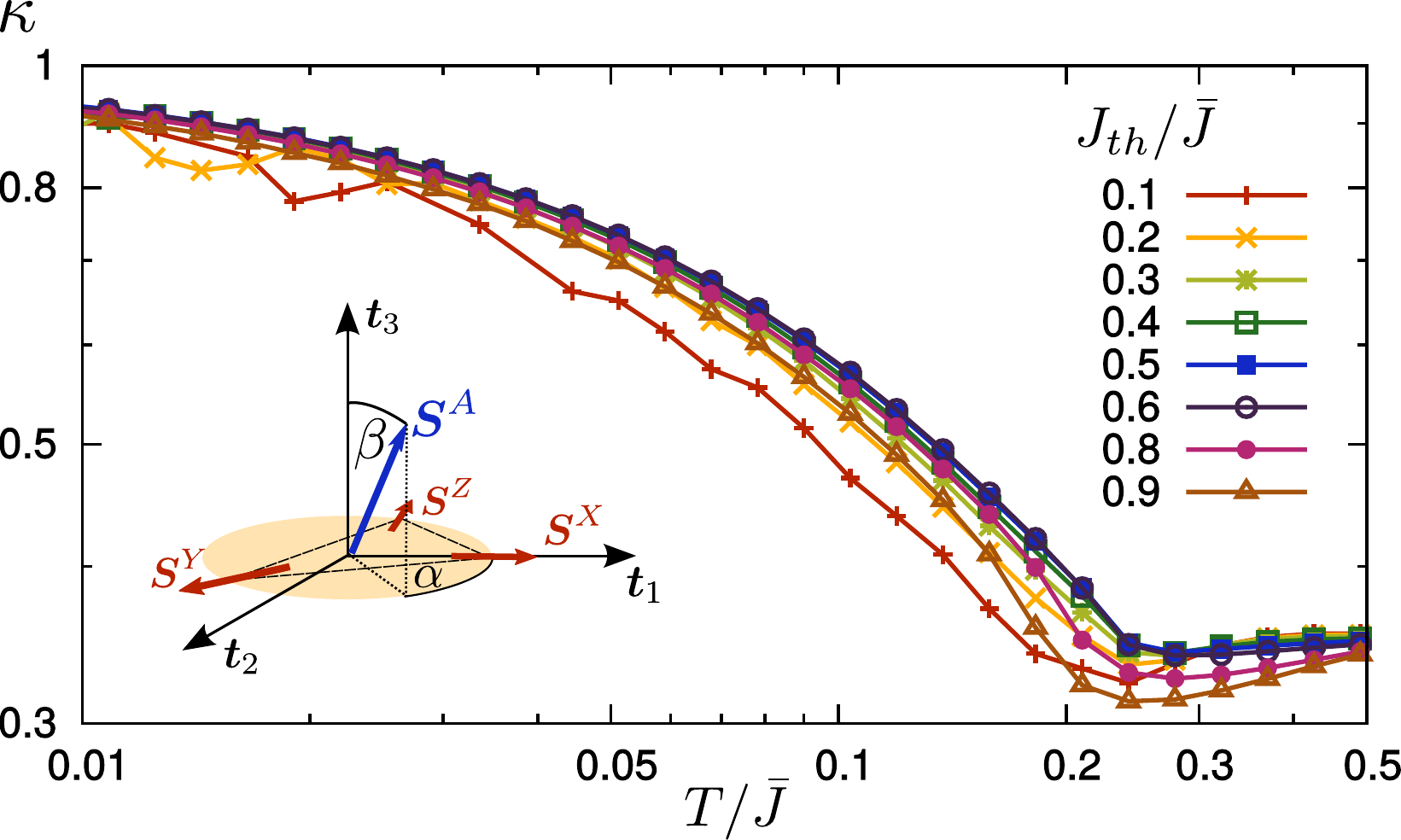}
\caption{(Color online) Coplanarity estimator $\kappa$ as a function of temperature for various values of $J_{th}/\bar{J}$ for system size $L = 60$. Inset shows definition of relative angles $\alpha$ and $\beta$. }
\label{fig:2}
\end{figure}

The emergent phases we are interested in occur for $J_{th} \leq \bar{J}$ where the zero temperature ground state is characterized by coplanar $120^\circ$ order of the triangular spins and N\'eel order of the honeycomb spins (see Fig.~\ref{fig:1})~\cite{PhysRevB.90.144435}. This order has $\text{SO(3)} \times \text{O(3)/O(2)}$ symmetry and is described by five Euler angles $(\theta, \phi, \psi) \times (\alpha, \beta)$. As shown in the inset of Fig.~\ref{fig:2}, the angles $(\alpha, \beta)$ describe the orientation of the honeycomb spins relative to the coordinate system $\bft_\gamma$ ($\gamma = 1,2,3$) set by the triangular spins. The Euler angles $(\theta, \phi, \psi)$ relate $\bft_\gamma$ to a fixed coordinate system. While the relative orientation can be changed without energy cost at $T=0$, thermal fluctuations induce order-by-disorder potentials~\cite{villain-JPhysFrance-1977,Shender82,PhysRevLett.62.2056}. Considering Gaussian thermal fluctuations around the classical ground state, one finds a contribution to the free energy equal to~\cite{PhysRevLett.68.855, Supplemental-MCClock}
\begin{align}
  \label{eq:2}
  \frac{F_{pot}}{N T} &= \cos(2 \beta) \Bigl[ 0.131 \frac{J_{th}^2}{\bar{J}^2} - 10^{-4} \frac{J_{th}^6}{\bar{J}^6}  \cos^2(3 \alpha) \Bigr] \,.
\end{align}
The first term forces the spins to become coplanar ($\beta = \pi/2$) below a coplanarity crossover temperature $T_{cp}$. More precisely, long-wavelength excitations out of the plane acquire a mass and are gapped out for $T < T_{cp}$. The second term shows that the remaining U(1) relative angle $\alpha$ is subject to a $\mathbb{Z}_6$ potential.

As shown in Fig.~\ref{fig:2}, we track this coplanarity crossover within the Monte-Carlo simulations by measuring the coplanarity estimator
\begin{align}
  \label{eq:1}
  \kappa &= 1 - \frac{3}{N} \sum_{j=1}^N \langle \cos^2 \beta_j \rangle \,,
\end{align}
where $\cos \beta_j = \bfss^A_j \cdot \bigl( \bfss^{t}_j \times \bfss^t_{j + \delta_{tt}}\bigr)$ with $\delta_{tt}$ being a nearest-neighbor vector on the triangular lattice. At high temperatures, where no relative spin configuration is preferred, one expects $\kappa = 1/3$, while for a completely coplanar state holds $\kappa = 1$. For uncorrelated $120^\circ$ and N\'eel order at $J_{th} = 0$ one finds $\kappa = 0$. Our Monte-Carlo results show that coplanarity develops as soon as $T \lesssim 0.25 \bar{J}$ and $\kappa$ smoothly approaches unity for lower temperatures. Interestingly, $\kappa$ depends only weakly on $J_{th}$ as long as $J_{th} \gtrsim \bar{J}/10$. We define the location of the coplanar crossover $T_{cp}$ shown in Fig.~\ref{fig:1} to be the location of the minimum of $\kappa$. Note that down to the lowest temperatures we observe substantial out-of-the plane fluctuations and $\kappa < 1$. We have identified these to be predominantly of short-wavelength nature.

Below the coplanar crossover temperature $T_{cp}$ one may define emergent XY spins $\bfm_{j}$ at all Bravais lattice sites via projecting the honeycomb spin $\bfss^A_j$ (or $\bfss^B_j \approx - \bfss^A_j$) onto the plane that is spanned by the three nearest-neighbor triangular spins and normalizing 
\begin{align}
  \label{eq:3}
  \bfm_{j} &= \frac{\bigl( \bfss^A_j \cdot \bft_{1,j}, \bfss^A_j \cdot \bft_{2,j} \bigr)}{\bigl\|\bigl( \bfss^A_j \cdot \bft_{1,j}, \bfss^A_j \cdot \bft_{2,j} \bigr) \bigr\|} = \bigl( \cos \alpha_j, \sin \alpha_j \bigr) \,.
\end{align}
We study the behavior of these emergent spins in the remainder of this paper. The local triangular triad $\bft_{\gamma,j}$ is defined as follows: the spins on the triangular lattice are first partitioned into three classes $\{\bfss^{t,X}_j, \bfss^{t,Y}_j, \bfss^{t,Z}_j\}$ as shown in Fig.~\ref{fig:2}. One then defines $\bft_{1,j} = \bfss^{t,X}_j$ and $\bft_{2,j}$ to point along the component of $\bfss^{t,Y}_j$ that is perpendicular to $\bft_{1,j}$. Finally, $\bft_{3,j} = \bft_{1,j} \times \bft_{2,j}$ completes the local triad. 
We show below that although the system exhibits out-of-the plane fluctuations and $\kappa < 1$, the emergent spins $\bfm_j$ decouple from these fluctuations and behave as U$(1)$ degrees of freedom. 

\begin{figure}[t!]
  \centering
  \includegraphics[width=\linewidth]{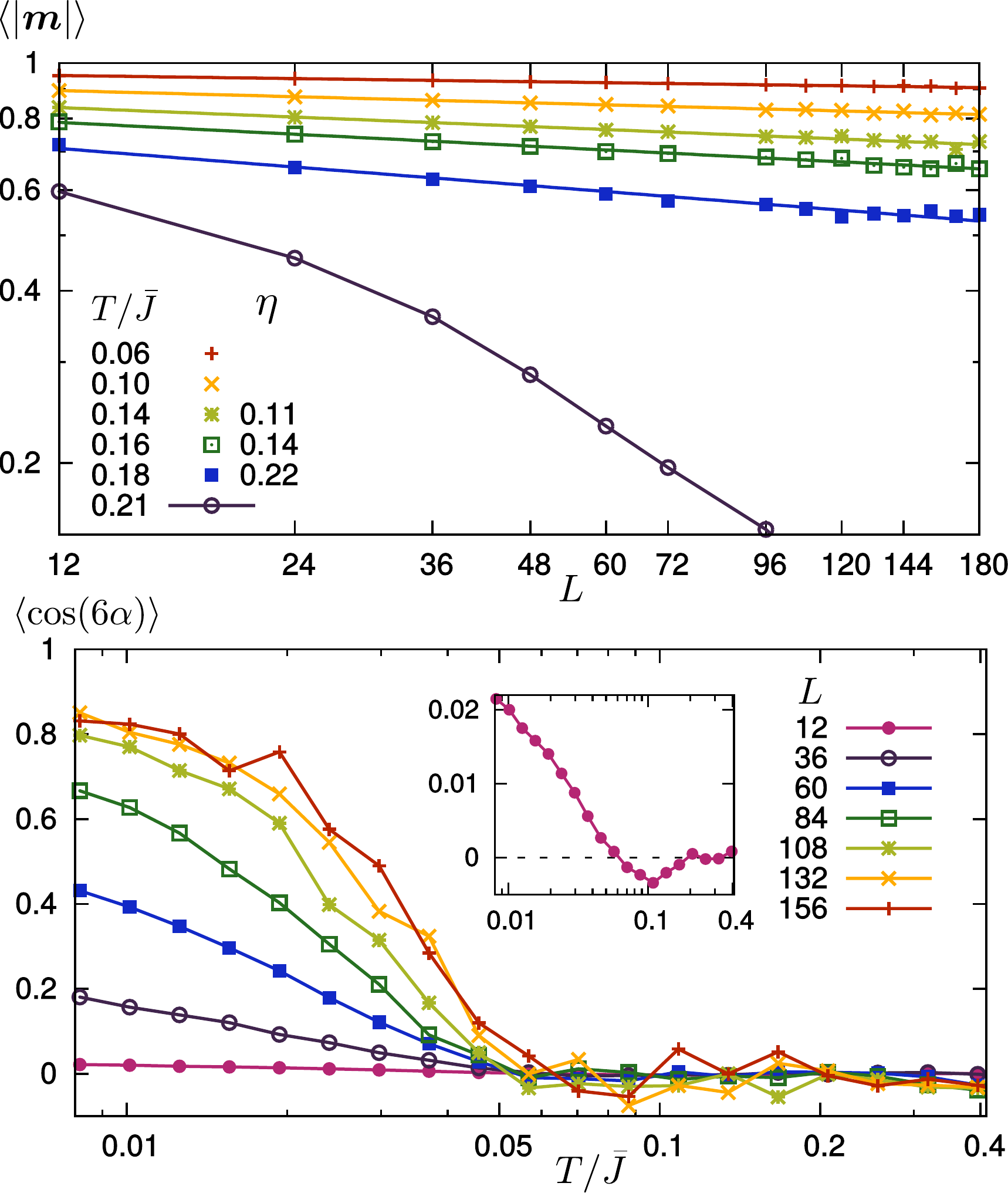}
  \caption{(color online). (a) XY magnetization amplitude $\av{|m|}$ as a function of linear system size $L$ for various temperatures $T/\bar{J}$ and fixed $J_{th}/\bar{J} = 0.8$. On a double logarithmic plot it exhibits linear scaling within the critical phase with indicated floating exponent $\eta(T)$. It bends down in the disordered phase. Due to the finite system size we cannot clearly observe a saturation (at a finite value) at low temperatures, but $\eta$ approaches zero in a linear fit. 
(b) Direction of the magnetization expressed as $\av{\cos (6 \alpha)}$ as a function of $T$ for $J_{th} = 0.9 \bar{J}$. A non-zero value signals breaking of the six-fold symmetry at low temperatures $T < T_<$. Inset shows $L=12$. }
  \label{fig:3}
\end{figure}

To map out the low temperature phase diagram we analyze the correlations of the emergent spins $\bfm_j$ in the following. First, we define the total magnetization as
\begin{align}
  \label{eq:8}
  \bfm = \frac{1}{N} \sum_{j=1}^N \bfm_j  = |\bfm| (\cos \alpha, \sin \alpha) \,.
\end{align}
The magnetization amplitude $|\bfm|$ depends on the (linear) system size $L$, in particular, it vanishes in the absence of long-range order for $L \rightarrow \infty$. Performing the Monte-Carlo average, we show the dependence of $\av{|\bfm|}$ with system size $L$ in Fig.~\ref{fig:3}(a). While it vanishes faster than algebraic at large temperatures, it exhibits power-law scaling $\av{|\bfm|} \propto L^{-\eta(T)/2}$ with $0 < \eta \lesssim 0.3$ for intermediate temperatures, a key signature of a critical phase. At the lowest temperatures, the exponent approaches zero and the magnetization saturates. To directly prove that the system develops (discrete) long-range order, we show the direction of the magnetization vector expressed as $\av{\cos(6 \alpha)}$ in Fig.~\ref{fig:3}(b). Clearly, $\av{\cos(6 \alpha)}$ approaches its saturation value of unity at low temperatures and large system sizes. The relative phase vector $\bfm$ points into one of the six directions preferred by the $\mathbb{Z}_6$ potential in Eq.~\eqref{eq:2}. The honeycomb spins are then aligned with one of the three triangular spin classes $\{S^{t,X}, S^{t,Y}, S^{t,Z}\}$, in agreement with the general order-from-disorder mechanism that spins tend to align their fluctuation Weiss fields to maximize their coupling~\cite{PhysRevLett.62.2056}.

\begin{figure}[t!]
\includegraphics[width=\linewidth]{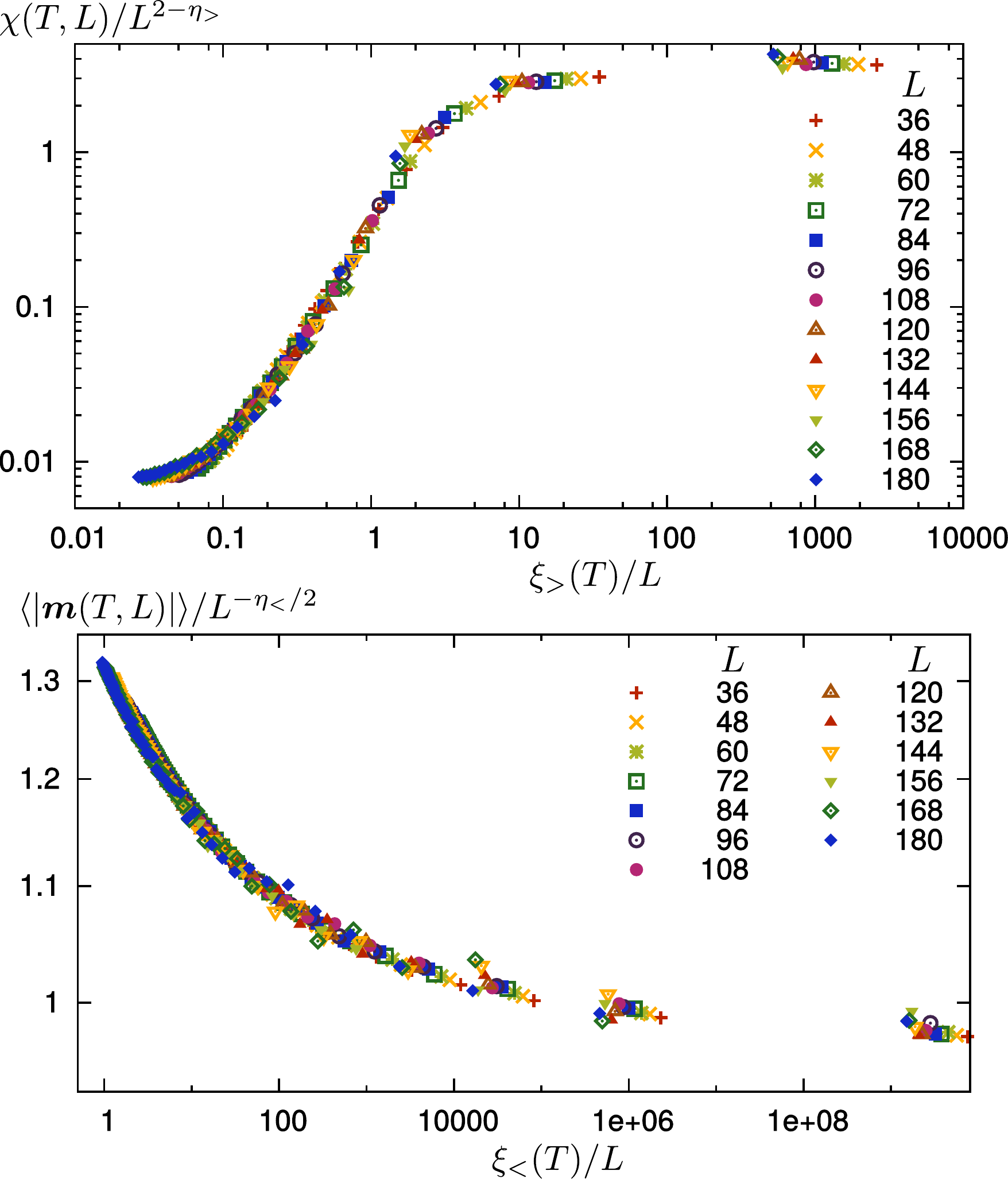} \caption{(Color online) Finite size scaling of susceptibility $\chi(T, L) = L^{2-\eta_>} Y_\chi(\xi_>/L)$ as a function of $\xi_>/L$ and magnetization $\av{|\bfm|(T, L)} = L^{-\eta_</2} Y_m(\xi_</L)$ as a function of $\xi_</L$ for $J_{th} = 0.6 \bar{J}$, $J_{tt} = 1.0$ and $\bar{J} = 1.22$. Best data collapse is obtained with a BKT scaling ansatz and yields $T_{<,>}$, $a_{<,>}$ and $\eta_{<,>}$ as given in the text. }
\label{fig:4}
\end{figure}

To determine the universality class of the phase transition and the transition temperatures $T_>$ and $T_<$, which partition the regimes of algebraic and long-range ordering, we perform a finite-size scaling analysis of the XY susceptibility and magnetization for various values of $J_{th}/\bar{J}$. As shown in Fig.~\ref{fig:4} we obtain perfect data collapse using a BKT scaling ansatz. Since the susceptibility diverges when the system enters a critical phase, we can detect the upper transition at $T_>$ by analyzing  
\begin{align}
  \label{eq:4}
  \chi(T, L) &= \frac{N}{T} \bigl\langle |\bfm|^2 \bigr \rangle = \frac{1}{N T} \Bigl\langle \bigl| \sum_j \bfm_{j} \bigr|^2 \Bigr\rangle 
\end{align}
for different temperatures $T$ and system sizes $L$. We employ a BKT ansatz for the correlation length $\xi_> = \exp\bigl(a_> \sqrt{T_>}/\sqrt{T - T_>}\bigr)$ with $a_>$ being a non-universal constant~\cite{PhysRevB.33.437, PhysRevB.68.104409, PhysRevLett.109.187201, PhysRevB.88.024410}. Since $\chi(T,\infty) \sim \xi_>(T)^{2-\eta_>}$ in the infinite system, it holds that $\chi(T,L) =  L^{2-\eta} Y_\chi(\xi_>(T)/L)$ with a universal function $Y_\chi(x)$. For $J_{th} = 0.6 \bar{J}$ we extract the values $T_> = 0.200(4) \bar{J}$, $a_> = 1.9(3)$ and $\eta_>= 0.25(1)$ from optimizing the collapse. This agrees very well with the theoretically expected value $\eta_> = 1/4$~\cite{PhysRevB.16.1217}. 

Performing the analysis for other values of $J_{th}$ yields data collapse of similar quality with a value $\eta_> = 0.25$ within error bars. This determines $T_>(\text{scal.})$ and the upper phase transition line in Fig.~\ref{fig:1}. As an independent way to determine $T_>$, we use the power-law scaling of the magnetization with the system size $L$, which is expected to be $\av{|\bfm|} \propto L^{-\eta/2}$ with $\eta = 1/4$ at the upper transition. This yields $T_>(\eta)$ included in Fig.~\ref{fig:1}. The two temperatures agree within error bars with $T_>(\eta)$ being systematically slightly larger. Finally, we note that we have also tried to achieve data collapse using a scaling ansatz corresponding to a second order phase transition, but the resulting collapse is worse in this case, especially for data points close to the phase transition. 

To determine the lower transition temperature $T_<$ we perform a finite size scaling analysis of the magnetization amplitude $\av{|\bfm|(T,L)}$. 
Since it holds in the infinite system that $\av{|\bfm|(T)} \propto \xi(T)_<^{-\eta_</2}$ with correlation length $\xi_< = \exp\bigl(a_< \sqrt{T_<}/\sqrt{T_< - T}\bigr)$ and non-universal factor $a_<$, it follows for a finite system that $\av{|\bfm|(T, L)} = L^{-\eta_</2} Y_m(\xi_<(T)/L)$, where $Y_m(x)$ is a universal function. In Fig.~\ref{fig:4}(b) we show the best data collapse for $J_{th} = 0.6 \bar{J}$ which yields $T_< = 0.18(1)$, $\eta_< = 0.11(1)$ and $a_<=5.0(5)$. This is in good agreement with the theoretically expected value of $\eta_< = 1/9$ at the lower transition~\cite{PhysRevB.16.1217,Cardy80}. 

Two independent ways to obtain $T_<$ are (i) to investigate the power-law scaling of $\av{|\bfm|}$ with system size and (ii) to directly look for the symmetry breaking as indicated by the quantity $\av{\cos(6 \alpha)}$. Using the first method, we find that our data can be fitted to $\log \av{|\bfm|} \propto -\frac{\eta(T)}{2} \log L$ with a temperature-dependent slope $\eta(T)$ that is monotonically decreasing over the full range $0 < T < T_>$. At high temperatures, we find $\eta(T_>) \approx 0.25$ (as expected) and we define $T_<(\eta)$ as the temperature where $\eta(T_<) = 1/9$. The fact that the system appears to be critical within our simulation even for lower temperatures (with an exponent $\eta < 1/9$) is a simple consequence of the fact that the system size is much smaller than the correlation length~\cite{PhysRevB.68.104409}. If we were able to reach larger system sizes in the simulation, we would eventually see a saturation of $\av{|\bfm|}$ to a finite value. 

Next we discuss the second method to detect $T_<$, namely direct observation of symmetry-breaking.  We see in Fig.~\ref{fig:3}(b) that $\av{\cos( 6 \alpha}$ approaches unity at low temperatures and large system sizes. In a finite-size system, we can observe this ordering only for not too small values of $J_{th} \geq 0.8 \bar{J}$ because the bare value of the order-from-disorder six-fold potential scales with $(J_{th}/\bar{J})^6$ with an additional small numerical prefactor $10^{-4}$ (see Eq.~\eqref{eq:2}). While the lower phase transition occurs when this potential becomes relevant at long lengthscales, independently of the bare value, the finite system size serves as a cut-off of the scaling making an effect of the potential only visible at sufficiently large bare values. To extract the transition temperature $T_<$ from $\av{\cos(6 \alpha)}$ we have to take into account that while at low temperatures the Gaussian order-from-disorder potential predicts free energy minima at $\alpha = 2 \pi n/6$ (in agreement with our simulation), at intermediate temperatures we observe in the finite size system a tendency of the spins to prefer a relative direction corresponding to a negative value of $\av{\cos(6 \alpha)}$ (see inset in Fig.~\ref{fig:3}(b)). This is presumably a result of nonlinear spin fluctuations around the classical ground state order, similarly to the effect of quenched disorder~\cite{PhysRevLett.62.2056}. We thus identify the transition temperature $T_<(\mathbb{Z}_6)$ as the location of the minimum of $\av{\cos(6 \alpha)}(T)$ which yields temperatures that are within error bars in agreement with the ones predicted from scaling.

We note that in the critical phase that develops for $T\in
[T_{<},T_{>}]$, the phase $\alpha $ behaves as a perfect, decoupled
XY order parameter.  Once the vortices bind at the BKT transition
$T_{>}$,  the ensemble of thermodynamically accessible states
divides up into distinct degenerate subspaces, each defined by a pair
of winding numbers $\{n_{x},n_{y} \}$ with  
\begin{equation}\label{}
n_{l}= \int_{0}^{L}\frac{dx_{l}}{2\pi} \nabla_{l}\alpha (x), \qquad \qquad (l=x,y),
\end{equation}
where $L$ is the linear size of the system, 
indicating the presence of an emergent topological phase~\cite{KosterlitzThouless-JPhysC-1973}. The multiple degeneracies of this state confirm the Polyakov hypothesis that a power-law phase is possible with a degenerate vacuum. 

In conclusion, employing extensive parallel-tempering Monte-Carlo simulations, 
we have presented conclusive evidence for an emergent critical phase in a 2D isotropic classical Heisenberg spin model at finite temperatures thereby realizing the Polyakov conjecture~\cite{polyakov_nlsm_75}. Using finite size scaling we have shown that the transitions are in the Berezinskii-Kosterlitz-Thouless universality class. At low temperatures, we find direct evidence of long-range order in the relative orientation of the spins via breaking of a discrete six-fold symmetry induced by an order-from-disorder potential. Direct numerical analysis of the spin stiffness tensor, the metric of the associated SO(3) $\times$ U(1) topological manifold, and its Ricci flow will be the topic of future work.

\acknowledgments
We acknowledge helpful discussions with B. Altshuler, K. Damle, J. Schmalian, M. D. Schulz, S. Trebst, and C. Weber. The Young Investigator Group of P.P.O. received financial support from the ``Concept for the Future'' of the Karlsruhe Institute of Technology (KIT) within the framework of the German Excellence Initiative. This work was supported by DOE grant DE-FG02-99ER45790 (P. Coleman) and P.C. and P.C. acknowledge visitors' support from the Institute for the Theory of Condensed Matter, Karlsruhe Institute of Technology (KIT). This work was carried out using the computational resource bwUniCluster funded by the Ministry of Science, Research and Arts and the Universities of the State of Baden-W\"urttemberg, Germany, within the framework program bwHPC.


\begin{thebibliography}{28}%
\makeatletter
\providecommand \@ifxundefined [1]{%
 \@ifx{#1\undefined}
}%
\providecommand \@ifnum [1]{%
 \ifnum #1\expandafter \@firstoftwo
 \else \expandafter \@secondoftwo
 \fi
}%
\providecommand \@ifx [1]{%
 \ifx #1\expandafter \@firstoftwo
 \else \expandafter \@secondoftwo
 \fi
}%
\providecommand \natexlab [1]{#1}%
\providecommand \enquote  [1]{``#1''}%
\providecommand \bibnamefont  [1]{#1}%
\providecommand \bibfnamefont [1]{#1}%
\providecommand \citenamefont [1]{#1}%
\providecommand \href@noop [0]{\@secondoftwo}%
\providecommand \href [0]{\begingroup \@sanitize@url \@href}%
\providecommand \@href[1]{\@@startlink{#1}\@@href}%
\providecommand \@@href[1]{\endgroup#1\@@endlink}%
\providecommand \@sanitize@url [0]{\catcode `\\12\catcode `\$12\catcode
  `\&12\catcode `\#12\catcode `\^12\catcode `\_12\catcode `\%12\relax}%
\providecommand \@@startlink[1]{}%
\providecommand \@@endlink[0]{}%
\providecommand \url  [0]{\begingroup\@sanitize@url \@url }%
\providecommand \@url [1]{\endgroup\@href {#1}{\urlprefix }}%
\providecommand \urlprefix  [0]{URL }%
\providecommand \Eprint [0]{\href }%
\providecommand \doibase [0]{http://dx.doi.org/}%
\providecommand \selectlanguage [0]{\@gobble}%
\providecommand \bibinfo  [0]{\@secondoftwo}%
\providecommand \bibfield  [0]{\@secondoftwo}%
\providecommand \translation [1]{[#1]}%
\providecommand \BibitemOpen [0]{}%
\providecommand \bibitemStop [0]{}%
\providecommand \bibitemNoStop [0]{.\EOS\space}%
\providecommand \EOS [0]{\spacefactor3000\relax}%
\providecommand \BibitemShut  [1]{\csname bibitem#1\endcsname}%
\let\auto@bib@innerbib\@empty
\bibitem [{\citenamefont
  {Polyakov}(1987)}]{Polyakov_GaugeFieldsAndStrings_Book}%
  \BibitemOpen
  \bibfield  {author} {\bibinfo {author} {\bibfnamefont {A.~M.}\ \bibnamefont
  {Polyakov}},\ }\href@noop {} {\emph {\bibinfo {title} {Gauge fields and
  strings}}},\ \bibinfo {series} {Contemporary concepts in physics},
  Vol.~\bibinfo {volume} {3}\ (\bibinfo  {publisher} {Harwood Academic
  Publishers, Chur, Switzerland},\ \bibinfo {year} {1987})\BibitemShut
  {NoStop}%
\bibitem [{\citenamefont {{Zinn-Justin}}(2002)}]{zinn-justin_QFT}%
  \BibitemOpen
  \bibfield  {author} {\bibinfo {author} {\bibfnamefont {J.}~\bibnamefont
  {{Zinn-Justin}}},\ }\href@noop {} {\emph {\bibinfo {title} {{Q}uantum {F}ield
  {T}heory and {C}ritical {P}henomena}}}\ (\bibinfo  {publisher} {Oxford
  University Press},\ \bibinfo {address} {New York, NY, USA},\ \bibinfo {year}
  {2002})\BibitemShut {NoStop}%
\bibitem [{\citenamefont {{P}olyakov}(1975)}]{polyakov_nlsm_75}%
  \BibitemOpen
  \bibfield  {author} {\bibinfo {author} {\bibfnamefont {A.~M.}\ \bibnamefont
  {{P}olyakov}},\ }\href@noop {} {\bibfield  {journal} {\bibinfo  {journal}
  {Phys. Lett. B}\ }\textbf {\bibinfo {volume} {59}},\ \bibinfo {pages} {79}
  (\bibinfo {year} {1975})}\BibitemShut {NoStop}%
\bibitem [{\citenamefont {Orth}\ \emph {et~al.}(2012)\citenamefont {Orth},
  \citenamefont {Chandra}, \citenamefont {Coleman},\ and\ \citenamefont
  {Schmalian}}]{PhysRevLett.109.237205}%
  \BibitemOpen
  \bibfield  {author} {\bibinfo {author} {\bibfnamefont {P.~P.}\ \bibnamefont
  {Orth}}, \bibinfo {author} {\bibfnamefont {P.}~\bibnamefont {Chandra}},
  \bibinfo {author} {\bibfnamefont {P.}~\bibnamefont {Coleman}}, \ and\
  \bibinfo {author} {\bibfnamefont {J.}~\bibnamefont {Schmalian}},\ }\href
  {\doibase 10.1103/PhysRevLett.109.237205} {\bibfield  {journal} {\bibinfo
  {journal} {Phys. Rev. Lett.}\ }\textbf {\bibinfo {volume} {109}},\ \bibinfo
  {pages} {237205} (\bibinfo {year} {2012})}\BibitemShut {NoStop}%
\bibitem [{\citenamefont {Orth}\ \emph {et~al.}(2014)\citenamefont {Orth},
  \citenamefont {Chandra}, \citenamefont {Coleman},\ and\ \citenamefont
  {Schmalian}}]{PhysRevB.89.094417}%
  \BibitemOpen
  \bibfield  {author} {\bibinfo {author} {\bibfnamefont {P.~P.}\ \bibnamefont
  {Orth}}, \bibinfo {author} {\bibfnamefont {P.}~\bibnamefont {Chandra}},
  \bibinfo {author} {\bibfnamefont {P.}~\bibnamefont {Coleman}}, \ and\
  \bibinfo {author} {\bibfnamefont {J.}~\bibnamefont {Schmalian}},\ }\href
  {\doibase 10.1103/PhysRevB.89.094417} {\bibfield  {journal} {\bibinfo
  {journal} {Phys. Rev. B}\ }\textbf {\bibinfo {volume} {89}},\ \bibinfo
  {pages} {094417} (\bibinfo {year} {2014})}\BibitemShut {NoStop}%
\bibitem [{\citenamefont {Cardy}(1980)}]{Cardy80}%
  \BibitemOpen
  \bibfield  {author} {\bibinfo {author} {\bibfnamefont {J.}~\bibnamefont
  {Cardy}},\ }\href@noop {} {\bibfield  {journal} {\bibinfo  {journal} {J.
  Phys. A: Math. Gen.}\ }\textbf {\bibinfo {volume} {13}},\ \bibinfo {pages}
  {1507} (\bibinfo {year} {1980})}\BibitemShut {NoStop}%
\bibitem [{\citenamefont {Chern}\ \emph {et~al.}(2011)\citenamefont {Chern},
  \citenamefont {Mellado},\ and\ \citenamefont
  {Tchernyshyov}}]{PhysRevLett.106.207202}%
  \BibitemOpen
  \bibfield  {author} {\bibinfo {author} {\bibfnamefont {G.-W.}\ \bibnamefont
  {Chern}}, \bibinfo {author} {\bibfnamefont {P.}~\bibnamefont {Mellado}}, \
  and\ \bibinfo {author} {\bibfnamefont {O.}~\bibnamefont {Tchernyshyov}},\
  }\href {\doibase 10.1103/PhysRevLett.106.207202} {\bibfield  {journal}
  {\bibinfo  {journal} {Phys. Rev. Lett.}\ }\textbf {\bibinfo {volume} {106}},\
  \bibinfo {pages} {207202} (\bibinfo {year} {2011})}\BibitemShut {NoStop}%
\bibitem [{\citenamefont {Chern}\ and\ \citenamefont
  {Tchernyshyov}(2012)}]{Chern12}%
  \BibitemOpen
  \bibfield  {author} {\bibinfo {author} {\bibfnamefont {G.-W.}\ \bibnamefont
  {Chern}}\ and\ \bibinfo {author} {\bibfnamefont {O.}~\bibnamefont
  {Tchernyshyov}},\ }\href@noop {} {\bibfield  {journal} {\bibinfo  {journal}
  {Phil. Trans. R. Soc. A}\ }\textbf {\bibinfo {volume} {370}},\ \bibinfo
  {pages} {5718} (\bibinfo {year} {2012})}\BibitemShut {NoStop}%
\bibitem [{\citenamefont {{C}handra}\ \emph {et~al.}(1990)\citenamefont
  {{C}handra}, \citenamefont {{C}oleman},\ and\ \citenamefont
  {{L}arkin}}]{PhysRevLett.64.88}%
  \BibitemOpen
  \bibfield  {author} {\bibinfo {author} {\bibfnamefont {P.}~\bibnamefont
  {{C}handra}}, \bibinfo {author} {\bibfnamefont {P.}~\bibnamefont
  {{C}oleman}}, \ and\ \bibinfo {author} {\bibfnamefont {A.~I.}\ \bibnamefont
  {{L}arkin}},\ }\href {\doibase 10.1103/PhysRevLett.64.88} {\bibfield
  {journal} {\bibinfo  {journal} {Phys. Rev. Lett.}\ }\textbf {\bibinfo
  {volume} {64}},\ \bibinfo {pages} {88} (\bibinfo {year} {1990})}\BibitemShut
  {NoStop}%
\bibitem [{\citenamefont {{F}ernandes}\ \emph {et~al.}(2010)\citenamefont
  {{F}ernandes}, \citenamefont {{V}an{B}ebber}, \citenamefont {{B}hattacharya},
  \citenamefont {{C}handra}, \citenamefont {{K}eppens}, \citenamefont
  {{M}andrus}, \citenamefont {{M}c{G}uire}, \citenamefont {{S}ales},
  \citenamefont {{S}efat},\ and\ \citenamefont
  {{S}chmalian}}]{PhysRevLett.105.157003}%
  \BibitemOpen
  \bibfield  {author} {\bibinfo {author} {\bibfnamefont {R.~M.}\ \bibnamefont
  {{F}ernandes}}, \bibinfo {author} {\bibfnamefont {L.~H.}\ \bibnamefont
  {{V}an{B}ebber}}, \bibinfo {author} {\bibfnamefont {S.}~\bibnamefont
  {{B}hattacharya}}, \bibinfo {author} {\bibfnamefont {P.}~\bibnamefont
  {{C}handra}}, \bibinfo {author} {\bibfnamefont {V.}~\bibnamefont
  {{K}eppens}}, \bibinfo {author} {\bibfnamefont {D.}~\bibnamefont
  {{M}andrus}}, \bibinfo {author} {\bibfnamefont {M.~A.}\ \bibnamefont
  {{M}c{G}uire}}, \bibinfo {author} {\bibfnamefont {B.~C.}\ \bibnamefont
  {{S}ales}}, \bibinfo {author} {\bibfnamefont {A.~S.}\ \bibnamefont
  {{S}efat}}, \ and\ \bibinfo {author} {\bibfnamefont {J.}~\bibnamefont
  {{S}chmalian}},\ }\href {\doibase 10.1103/PhysRevLett.105.157003} {\bibfield
  {journal} {\bibinfo  {journal} {Phys. Rev. Lett.}\ }\textbf {\bibinfo
  {volume} {105}},\ \bibinfo {pages} {157003} (\bibinfo {year}
  {2010})}\BibitemShut {NoStop}%
\bibitem [{\citenamefont {{F}ernandes}\ \emph {et~al.}(2012)\citenamefont
  {{F}ernandes}, \citenamefont {{C}hubukov}, \citenamefont {{K}nolle},
  \citenamefont {{E}remin},\ and\ \citenamefont
  {{S}chmalian}}]{PhysRevB.85.024534}%
  \BibitemOpen
  \bibfield  {author} {\bibinfo {author} {\bibfnamefont {R.~M.}\ \bibnamefont
  {{F}ernandes}}, \bibinfo {author} {\bibfnamefont {A.~V.}\ \bibnamefont
  {{C}hubukov}}, \bibinfo {author} {\bibfnamefont {J.}~\bibnamefont
  {{K}nolle}}, \bibinfo {author} {\bibfnamefont {I.}~\bibnamefont {{E}remin}},
  \ and\ \bibinfo {author} {\bibfnamefont {J.}~\bibnamefont {{S}chmalian}},\
  }\href {\doibase 10.1103/PhysRevB.85.024534} {\bibfield  {journal} {\bibinfo
  {journal} {Phys. Rev. B}\ }\textbf {\bibinfo {volume} {85}},\ \bibinfo
  {pages} {024534} (\bibinfo {year} {2012})}\BibitemShut {NoStop}%
\bibitem [{\citenamefont {{W}eber}\ \emph {et~al.}(2003)\citenamefont
  {{W}eber}, \citenamefont {{C}apriotti}, \citenamefont {{M}isguich},
  \citenamefont {{B}ecca}, \citenamefont {{E}lhajal},\ and\ \citenamefont
  {{M}ila}}]{PhysRevLett.91.177202}%
  \BibitemOpen
  \bibfield  {author} {\bibinfo {author} {\bibfnamefont {C.}~\bibnamefont
  {{W}eber}}, \bibinfo {author} {\bibfnamefont {L.}~\bibnamefont
  {{C}apriotti}}, \bibinfo {author} {\bibfnamefont {G.}~\bibnamefont
  {{M}isguich}}, \bibinfo {author} {\bibfnamefont {F.}~\bibnamefont {{B}ecca}},
  \bibinfo {author} {\bibfnamefont {M.}~\bibnamefont {{E}lhajal}}, \ and\
  \bibinfo {author} {\bibfnamefont {F.}~\bibnamefont {{M}ila}},\ }\href
  {\doibase 10.1103/PhysRevLett.91.177202} {\bibfield  {journal} {\bibinfo
  {journal} {Phys. Rev. Lett.}\ }\textbf {\bibinfo {volume} {91}},\ \bibinfo
  {pages} {177202} (\bibinfo {year} {2003})}\BibitemShut {NoStop}%
\bibitem [{\citenamefont {{C}apriotti}\ \emph {et~al.}(2004)\citenamefont
  {{C}apriotti}, \citenamefont {{F}ubini}, \citenamefont {{R}oscilde},\ and\
  \citenamefont {{T}ognetti}}]{PhysRevLett.92.157202}%
  \BibitemOpen
  \bibfield  {author} {\bibinfo {author} {\bibfnamefont {L.}~\bibnamefont
  {{C}apriotti}}, \bibinfo {author} {\bibfnamefont {A.}~\bibnamefont
  {{F}ubini}}, \bibinfo {author} {\bibfnamefont {T.}~\bibnamefont
  {{R}oscilde}}, \ and\ \bibinfo {author} {\bibfnamefont {V.}~\bibnamefont
  {{T}ognetti}},\ }\href {\doibase 10.1103/PhysRevLett.92.157202} {\bibfield
  {journal} {\bibinfo  {journal} {Phys. Rev. Lett.}\ }\textbf {\bibinfo
  {volume} {92}},\ \bibinfo {pages} {157202} (\bibinfo {year}
  {2004})}\BibitemShut {NoStop}%
\bibitem [{\citenamefont {{J}os\'e}\ \emph {et~al.}(1977)\citenamefont
  {{J}os\'e}, \citenamefont {{K}adanoff}, \citenamefont {{K}irkpatrick},\ and\
  \citenamefont {{N}elson}}]{PhysRevB.16.1217}%
  \BibitemOpen
  \bibfield  {author} {\bibinfo {author} {\bibfnamefont {J.~V.}\ \bibnamefont
  {{J}os\'e}}, \bibinfo {author} {\bibfnamefont {L.~P.}\ \bibnamefont
  {{K}adanoff}}, \bibinfo {author} {\bibfnamefont {S.}~\bibnamefont
  {{K}irkpatrick}}, \ and\ \bibinfo {author} {\bibfnamefont {D.~R.}\
  \bibnamefont {{N}elson}},\ }\href {\doibase 10.1103/PhysRevB.16.1217}
  {\bibfield  {journal} {\bibinfo  {journal} {Phys. Rev. B}\ }\textbf {\bibinfo
  {volume} {16}},\ \bibinfo {pages} {1217} (\bibinfo {year}
  {1977})}\BibitemShut {NoStop}%
\bibitem [{\citenamefont {Miyatake}\ \emph {et~al.}(1986)\citenamefont
  {Miyatake}, \citenamefont {Yamamoto}, \citenamefont {Kim}, \citenamefont
  {Toyonaga},\ and\ \citenamefont {Nagai}}]{0022-3719-19-14-020}%
  \BibitemOpen
  \bibfield  {author} {\bibinfo {author} {\bibfnamefont {Y.}~\bibnamefont
  {Miyatake}}, \bibinfo {author} {\bibfnamefont {M.}~\bibnamefont {Yamamoto}},
  \bibinfo {author} {\bibfnamefont {J.~J.}\ \bibnamefont {Kim}}, \bibinfo
  {author} {\bibfnamefont {M.}~\bibnamefont {Toyonaga}}, \ and\ \bibinfo
  {author} {\bibfnamefont {O.}~\bibnamefont {Nagai}},\ }\href
  {http://stacks.iop.org/0022-3719/19/i=14/a=020} {\bibfield  {journal}
  {\bibinfo  {journal} {J. Phys. C}\ }\textbf {\bibinfo {volume} {19}},\
  \bibinfo {pages} {2539} (\bibinfo {year} {1986})}\BibitemShut {NoStop}%
\bibitem [{\citenamefont {Marinari}\ and\ \citenamefont
  {Parisi}(1992)}]{0295-5075-19-6-002}%
  \BibitemOpen
  \bibfield  {author} {\bibinfo {author} {\bibfnamefont {E.}~\bibnamefont
  {Marinari}}\ and\ \bibinfo {author} {\bibfnamefont {G.}~\bibnamefont
  {Parisi}},\ }\href {http://stacks.iop.org/0295-5075/19/i=6/a=002} {\bibfield
  {journal} {\bibinfo  {journal} {EPL}\ }\textbf {\bibinfo {volume} {19}},\
  \bibinfo {pages} {451} (\bibinfo {year} {1992})}\BibitemShut {NoStop}%
\bibitem [{\citenamefont {Hukushima}\ and\ \citenamefont
  {Nemoto}(1996)}]{doi:10.1143/JPSJ.65.1604}%
  \BibitemOpen
  \bibfield  {author} {\bibinfo {author} {\bibfnamefont {K.}~\bibnamefont
  {Hukushima}}\ and\ \bibinfo {author} {\bibfnamefont {K.}~\bibnamefont
  {Nemoto}},\ }\href {\doibase 10.1143/JPSJ.65.1604} {\bibfield  {journal}
  {\bibinfo  {journal} {J. Phys. Soc. Jpn.}\ }\textbf {\bibinfo {volume}
  {65}},\ \bibinfo {pages} {1604} (\bibinfo {year} {1996})}\BibitemShut
  {NoStop}%
\bibitem [{\citenamefont {Jeevanesan}\ and\ \citenamefont
  {Orth}(2014)}]{PhysRevB.90.144435}%
  \BibitemOpen
  \bibfield  {author} {\bibinfo {author} {\bibfnamefont {B.}~\bibnamefont
  {Jeevanesan}}\ and\ \bibinfo {author} {\bibfnamefont {P.~P.}\ \bibnamefont
  {Orth}},\ }\href {\doibase 10.1103/PhysRevB.90.144435} {\bibfield  {journal}
  {\bibinfo  {journal} {Phys. Rev. B}\ }\textbf {\bibinfo {volume} {90}},\
  \bibinfo {pages} {144435} (\bibinfo {year} {2014})}\BibitemShut {NoStop}%
\bibitem [{\citenamefont {{V}illain}(1977)}]{villain-JPhysFrance-1977}%
  \BibitemOpen
  \bibfield  {author} {\bibinfo {author} {\bibfnamefont {J.}~\bibnamefont
  {{V}illain}},\ }\href@noop {} {\bibfield  {journal} {\bibinfo  {journal} {J.
  Phys France}\ }\textbf {\bibinfo {volume} {38}},\ \bibinfo {pages} {385}
  (\bibinfo {year} {1977})}\BibitemShut {NoStop}%
\bibitem [{\citenamefont {{S}hender}(1982)}]{Shender82}%
  \BibitemOpen
  \bibfield  {author} {\bibinfo {author} {\bibfnamefont {E.}~\bibnamefont
  {{S}hender}},\ }\href@noop {} {\bibfield  {journal} {\bibinfo  {journal}
  {Sov. Phys. JETP}\ }\textbf {\bibinfo {volume} {56}},\ \bibinfo {pages} {178}
  (\bibinfo {year} {1982})}\BibitemShut {NoStop}%
\bibitem [{\citenamefont {{H}enley}(1989)}]{PhysRevLett.62.2056}%
  \BibitemOpen
  \bibfield  {author} {\bibinfo {author} {\bibfnamefont {C.~L.}\ \bibnamefont
  {{H}enley}},\ }\href {\doibase 10.1103/PhysRevLett.62.2056} {\bibfield
  {journal} {\bibinfo  {journal} {Phys. Rev. Lett.}\ }\textbf {\bibinfo
  {volume} {62}},\ \bibinfo {pages} {2056} (\bibinfo {year}
  {1989})}\BibitemShut {NoStop}%
\bibitem [{\citenamefont {{C}halker}\ \emph {et~al.}(1992)\citenamefont
  {{C}halker}, \citenamefont {{H}oldsworth},\ and\ \citenamefont
  {{S}hender}}]{PhysRevLett.68.855}%
  \BibitemOpen
  \bibfield  {author} {\bibinfo {author} {\bibfnamefont {J.~T.}\ \bibnamefont
  {{C}halker}}, \bibinfo {author} {\bibfnamefont {P.~C.~W.}\ \bibnamefont
  {{H}oldsworth}}, \ and\ \bibinfo {author} {\bibfnamefont {E.~F.}\
  \bibnamefont {{S}hender}},\ }\href {\doibase 10.1103/PhysRevLett.68.855}
  {\bibfield  {journal} {\bibinfo  {journal} {Phys. Rev. Lett.}\ }\textbf
  {\bibinfo {volume} {68}},\ \bibinfo {pages} {855} (\bibinfo {year}
  {1992})}\BibitemShut {NoStop}%
\bibitem [{Sup()}]{Supplemental-MCClock}%
  \BibitemOpen
  \href@noop {} {}\bibinfo {note} {The Supplemental Material contains details
  on the derivation of the order from disorder potential.}\BibitemShut {Stop}%
\bibitem [{\citenamefont {Challa}\ and\ \citenamefont
  {Landau}(1986)}]{PhysRevB.33.437}%
  \BibitemOpen
  \bibfield  {author} {\bibinfo {author} {\bibfnamefont {M.}~\bibnamefont
  {Challa}}\ and\ \bibinfo {author} {\bibfnamefont {D.}~\bibnamefont
  {Landau}},\ }\href {\doibase 10.1103/PhysRevB.33.437} {\bibfield  {journal}
  {\bibinfo  {journal} {Phys. Rev. B}\ }\textbf {\bibinfo {volume} {33}},\
  \bibinfo {pages} {437} (\bibinfo {year} {1986})}\BibitemShut {NoStop}%
\bibitem [{\citenamefont {{I}sakov}\ and\ \citenamefont
  {{M}oessner}(2003)}]{PhysRevB.68.104409}%
  \BibitemOpen
  \bibfield  {author} {\bibinfo {author} {\bibfnamefont {S.~V.}\ \bibnamefont
  {{I}sakov}}\ and\ \bibinfo {author} {\bibfnamefont {R.}~\bibnamefont
  {{M}oessner}},\ }\href {\doibase 10.1103/PhysRevB.68.104409} {\bibfield
  {journal} {\bibinfo  {journal} {Phys. Rev. B}\ }\textbf {\bibinfo {volume}
  {68}},\ \bibinfo {pages} {104409} (\bibinfo {year} {2003})}\BibitemShut
  {NoStop}%
\bibitem [{\citenamefont {{P}rice}\ and\ \citenamefont
  {{P}erkins}(2012)}]{PhysRevLett.109.187201}%
  \BibitemOpen
  \bibfield  {author} {\bibinfo {author} {\bibfnamefont {C.~C.}\ \bibnamefont
  {{P}rice}}\ and\ \bibinfo {author} {\bibfnamefont {N.~B.}\ \bibnamefont
  {{P}erkins}},\ }\href {\doibase 10.1103/PhysRevLett.109.187201} {\bibfield
  {journal} {\bibinfo  {journal} {Phys. Rev. Lett.}\ }\textbf {\bibinfo
  {volume} {109}},\ \bibinfo {pages} {187201} (\bibinfo {year}
  {2012})}\BibitemShut {NoStop}%
\bibitem [{\citenamefont {Price}\ and\ \citenamefont
  {Perkins}(2013)}]{PhysRevB.88.024410}%
  \BibitemOpen
  \bibfield  {author} {\bibinfo {author} {\bibfnamefont {C.}~\bibnamefont
  {Price}}\ and\ \bibinfo {author} {\bibfnamefont {N.~B.}\ \bibnamefont
  {Perkins}},\ }\href {\doibase 10.1103/PhysRevB.88.024410} {\bibfield
  {journal} {\bibinfo  {journal} {Phys. Rev. B}\ }\textbf {\bibinfo {volume}
  {88}},\ \bibinfo {pages} {024410} (\bibinfo {year} {2013})}\BibitemShut
  {NoStop}%
\bibitem [{\citenamefont {{K}osterlitz}\ and\ \citenamefont
  {{T}houless}(1973)}]{KosterlitzThouless-JPhysC-1973}%
  \BibitemOpen
  \bibfield  {author} {\bibinfo {author} {\bibfnamefont {J.~M.}\ \bibnamefont
  {{K}osterlitz}}\ and\ \bibinfo {author} {\bibfnamefont {D.~J.}\ \bibnamefont
  {{T}houless}},\ }\href {\doibase doi:10.1088/0022-3719/6/7/010} {\bibfield
  {journal} {\bibinfo  {journal} {J. Phys. C: Solid St. Phys.}\ }\textbf
  {\bibinfo {volume} {6}},\ \bibinfo {pages} {1181} (\bibinfo {year}
  {1973})}\BibitemShut {NoStop}%
\end{thebibliography}

%

\pagebreak


\section*{Supplementary material}

# Supplemental to "Emergent Power-Law Phase in the 2D Heisenberg Windmill Antiferromagnet: A Computational Experiment"

Bhilahari Jeevanesan,[1] Premala Chandra,[2] Piers Coleman,[2,3] and Peter P. Orth[1,4]

[1]*Institute for Theory of Condensed Matter, Karlsruhe Institute of Technology (KIT), 76131 Karlsruhe, Germany*
[2]*Center for Materials Theory, Rutgers University, Piscataway, New Jersey 08854, USA*
[3]*Hubbard Theory Consortium and Department of Physics,
Royal Holloway, University of London, Egham, Surrey TW20 0EX, UK*
[4]*School of Physics and Astronomy, University of Minnesota, Minneapolis, Minnesota 55455, USA*
(Dated: June 11, 2015)

**Analytic calculation of the order-by-disorder potential**

In this appendix we summarize our calculation of the order-by-disorder potential terms. The aim of this calculation is to determine the dependence of the free-energy on the manifold of ground-state configurations (the calculation follows [1–3],[4]). In order to carry out the calculation we parametrize the spins as small fluctuations around one of the ground states. We choose the plane of spins of the triangular lattice spins in their groundstate to lie in the $xy$ plane. The honeycomb spins have an order parameter, that is described by angles $\beta$ and $\alpha$. The fluctuations of the triangular sublattice spins are called $(\rho, \sigma, \tau)$ and those of the honeycomb spins are called $(\mu, \nu)$. We choose the following parametrization of the fluctuations

$$\boldsymbol{n}^A = \begin{pmatrix} 1 - \frac{\rho_y^2 + \rho_z^2}{2} \\ \rho_y \\ \rho_z \end{pmatrix}$$

$$\boldsymbol{n}^B = R \begin{pmatrix} 1 - \frac{\sigma_y^2 + \sigma_z^2}{2} \\ \sigma_y \\ \sigma_z \end{pmatrix} = \begin{pmatrix} -\frac{1}{2} - \frac{\sqrt{3}}{2}\sigma_y + \frac{\sigma_y^2 + \sigma_z^2}{4} \\ \frac{\sqrt{3}}{2} - \frac{1}{2}\sigma_y - \frac{\sqrt{3}}{4}(\sigma_y^2 + \sigma_z^2) \\ \sigma_z \end{pmatrix}$$

$$\boldsymbol{n}^C = R^2 \begin{pmatrix} 1 - \frac{\tau_y^2 + \tau_z^2}{2} \\ \tau_y \\ \tau_z \end{pmatrix} = \begin{pmatrix} -\frac{1}{2} + \frac{\sqrt{3}}{2}\tau_y + \frac{\tau_y^2 + \tau_z^2}{4} \\ -\frac{\sqrt{3}}{2} - \frac{1}{2}\tau_y + \frac{\sqrt{3}}{4}(\tau_y^2 + \tau_z^2) \\ \tau_z \end{pmatrix}$$

$$\boldsymbol{h}^A = \left(1 - \frac{\mu_1^2 + \mu_2^2}{2}\right)\boldsymbol{h}_0 + \mu_1\boldsymbol{\beta} + \mu_2\boldsymbol{\alpha}$$

$$\boldsymbol{h}^B = -\left[\left(1 - \frac{\nu_1^2 + \nu_2^2}{2}\right)\boldsymbol{h}_0 + \nu_1\boldsymbol{\beta} + \nu_2\boldsymbol{\alpha}\right]$$

with unit vectors

$$\boldsymbol{h}_0 = \begin{pmatrix} \sin\beta\cos\alpha \\ \sin\beta\sin\alpha \\ \cos\beta \end{pmatrix}, \quad \boldsymbol{\beta} = \begin{pmatrix} \cos\beta\cos\alpha \\ \cos\beta\sin\alpha \\ \sin\beta \end{pmatrix}, \quad \boldsymbol{\alpha} = \begin{pmatrix} -\sin\alpha \\ \cos\alpha \\ 0 \end{pmatrix}$$

and where $R = R_z(2\pi/3)$ is the rotation matrix that rotates spins around the z-axis by $2\pi/3$. With this parametrization it is a straightforward task to rewrite the Hamiltonian in the following form. The triangular spin Hamiltonian becomes

$$H_t = J_{tt} \sum_i \boldsymbol{n}_i^A \cdot \left(\sum_{j(i)} \boldsymbol{n}_j^B + \sum_{j(i)} \boldsymbol{n}_j^C\right) + J_{tt} \sum_i \boldsymbol{n}_i^B \cdot \sum_{j(i)} \boldsymbol{n}_j^C$$

here the notation $j(i)$ denotes summation over the neighbors $j$ of the spin with index $i$. This is

$$H_t = -\frac{3}{2}NJ_{tt} + \delta H_t$$

with

$$\delta H_t = J_{tt} \sum_{\langle i,j \rangle} \left[ \frac{1}{4} \left( \rho_{iy}^2 + \rho_{iz}^2 + \sigma_{jy}^2 + \sigma_{jz}^2 \right) - \frac{1}{2} \rho_{iy}\sigma_{jy} + \rho_{iz}\sigma_{jz} \right]$$
$$+ J_{tt} \sum_{\langle i,j \rangle} \left[ \frac{1}{4} \left( \rho_{iy}^2 + \rho_{iz}^2 + \tau_{jy}^2 + \tau_{jz}^2 \right) - \frac{1}{2} \rho_{iy}\tau_{jy} + \rho_{iz}\tau_{jz} \right]$$
$$+ J_{tt} \sum_{\langle i,j \rangle} \left[ \frac{1}{4} \left( \sigma_{iy}^2 + \sigma_{iz}^2 + \tau_{jy}^2 + \tau_{jz}^2 \right) - \frac{1}{2} \sigma_{iy}\tau_{jy} + \sigma_{iz}\tau_{jz} \right],$$

where all the linear terms have canceled out. Similarly we have

$$\delta H_h = J_{hh} \sum_i \frac{3}{2} \left[ \left(\mu_{i1}^A\right)^2 + \left(\mu_{i2}^A\right)^2 + \left(\nu_{i1}^A\right)^2 + \left(\nu_{i2}^A\right)^2 + \left(\mu_{i1}^B\right)^2 + \left(\mu_{i2}^B\right)^2 + \left(\nu_{i1}^B\right)^2 + \left(\nu_{i2}^B\right)^2 + \left(\mu_{i1}^C\right)^2 + \left(\mu_{i2}^C\right)^2 + \left(\nu_{i1}^C\right)^2 + \left(\nu_{i2}^C\right)^2 \right]$$
$$- J_{hh} \sum_i \mu_{i1}^A \left[ \nu_{i+\hat{1}+\hat{2},1}^A + \nu_{i1}^B + \nu_{i1}^C \right] + \mu_{i1}^B \left[ \nu_{i+\hat{1}+\hat{2},1}^A + \nu_{i+\hat{1}+\hat{2},1}^B + \nu_{i+\hat{1},1}^C \right] + \mu_{i1}^C \left[ \nu_{i+\hat{1}+\hat{2},1}^A + \nu_{i+\hat{2},1}^B + \nu_{i+\hat{1}+\hat{2},1}^C \right]$$
$$- J_{hh} \sum_i \mu_{i2}^A \left[ \nu_{i+\hat{1}+\hat{2},2}^A + \nu_{i2}^B + \nu_{i2}^C \right] + \mu_{i2}^B \left[ \nu_{i+\hat{1}+\hat{2},2}^A + \nu_{i+\hat{1}+\hat{2},2}^B + \nu_{i+\hat{1},2}^C \right] + \mu_{i2}^C \left[ \nu_{i+\hat{1}+\hat{2},2}^A + \nu_{i+\hat{2},2}^B + \nu_{i+\hat{1}+\hat{2},2}^C \right]$$

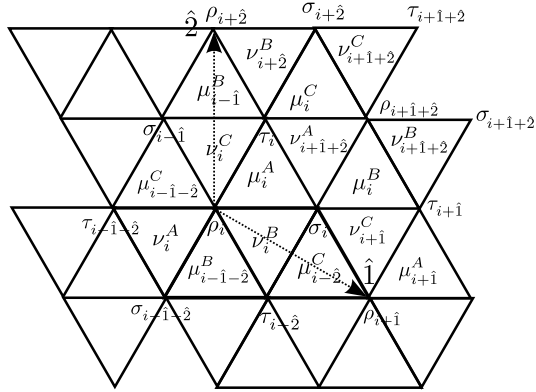

FIG. 1. Labelling of sites inside the enlarged magnetic unit-cell.

and finally

$$\begin{aligned}H_{th} &= J_{th}\sum_i \cos\alpha \rho_{iy}\left(\mu_{i2}^A + \mu_{i-\hat{1}-\hat{2},2}^B + \mu_{i-\hat{1}-\hat{2},2}^C\right) + \cos\beta\sin\alpha\rho_{iy}\left(\mu_{i1}^A + \mu_{i-\hat{1}-\hat{2},1}^B + \mu_{i-\hat{1}-\hat{2},1}^C\right)\\
&- \sin\beta\rho_{iz}\left(\mu_{i1}^A + \mu_{i-\hat{1}-\hat{2},1}^B + \mu_{i-\hat{1}-\hat{2},1}^C\right) - \cos\alpha\rho_{iy}\left(\nu_{i2}^A + \nu_{i2}^B + \nu_{i2}^C\right) - \cos\beta\sin\alpha\rho_{iy}\left(\nu_{i1}^A + \nu_{i1}^B + \nu_{i1}^C\right)\\
&+ \sin\beta\rho_{iz}\left(\nu_{i1}^A + \nu_{i1}^B + \nu_{i1}^C\right) + J_{th}\sum_i \frac{\sqrt{3}\sin\alpha - \cos\alpha}{2}\sigma_{iy}\left(\mu_{i2}^A + \mu_{i2}^B + \mu_{i-\hat{2},2}^C\right)\\
&- \frac{\sqrt{3}\cos\alpha + \sin\alpha}{2}\cos\beta\sigma_{iy}\left(\mu_{i1}^A + \mu_{i1}^B + \mu_{i-\hat{2},1}^C\right) - \sin\beta\sigma_{iz}\left(\mu_{i1}^B + \mu_{i1}^A + \mu_{i-\hat{2},1}^C\right)\\
&+ \frac{-\sqrt{3}\sin\alpha + \cos\alpha}{2}\sigma_{iy}\left(\nu_{i+\hat{1}+\hat{2},2}^A + \nu_{i2}^B + \nu_{i+\hat{1},2}^C\right) + \frac{\sqrt{3}\cos\alpha + \sin\alpha}{2}\cos\beta\sigma_{iy}\left(\nu_{i+\hat{1}+\hat{2},1}^A + \nu_{i1}^B + \nu_{i+\hat{1},1}^C\right)\\
&+ \sin\beta\sigma_{iz}\left(\nu_{i+\hat{1}+\hat{2},1}^A + \nu_{i1}^B + \nu_{i+\hat{1},1}^C\right) + J_{th}\sum_i -\frac{\sqrt{3}\sin\alpha + \cos\alpha}{2}\tau_{iy}\left(\mu_{i2}^A + \mu_{i-\hat{1},2}^B + \mu_{i2}^C\right)\\
&+ \frac{\sqrt{3}\cos\alpha - \sin\alpha}{2}\cos\beta\tau_{iy}\left(\mu_{i1}^A + \mu_{i-\hat{1},1}^B + \mu_{i1}^C\right) - \sin\beta\tau_{iz}\left(\mu_{i1}^A + \mu_{i-\hat{1},1}^B + \mu_{i1}^C\right)\\
&+ \frac{\sqrt{3}\sin\alpha + \cos\alpha}{2}\tau_{iy}\left(\nu_{i+\hat{1}+\hat{2},2}^A + \nu_{i+\hat{2},2}^B + \nu_{i2}^C\right) - \frac{\sqrt{3}\cos\alpha - \sin\alpha}{2}\cos\beta\tau_{iy}\left(\nu_{i+\hat{1}+\hat{2},1}^A + \nu_{i+\hat{2},1}^B + \nu_{i1}^C\right)\\
&+ \sin\beta\tau_{iz}\left(\nu_{i+\hat{1}+\hat{2},1}^A + \nu_{i+\hat{2},1}^B + \nu_{i1}^C\right).\end{aligned}$$

We transform this into fourier space with

$$\rho_i = \rho(\boldsymbol{r}_i) = \frac{1}{(2\pi)^2}\int d^2q\, e^{i\boldsymbol{q}\boldsymbol{r}_i}\rho(\boldsymbol{q})$$

and rewrite the resulting total Hamiltonian in matrix form. This matrix has dimensions $18\times 18$. It may, however, be written in extremely compact block form. This will facilitate the calculation of the free energy below. The full Hamiltonian is

$$\delta H = \frac{N}{6(2\pi)^2}\int d^2q\, \psi^\dagger \begin{pmatrix} J_{tt}M_t & J_{th}M_{th}^\dagger \\ J_{th}M_{th} & J_{hh}M_h \end{pmatrix}\psi.$$

with

$$M_t = \begin{pmatrix} \frac{15}{2}\mathbb{I} - \frac{j_0 j_0^\dagger}{2} & 0 \\ 0 & 3\mathbb{I} + j_0 j_0^\dagger \end{pmatrix}$$

$$M_h = \begin{pmatrix} 3\mathbb{I} & -e^{i(q_1+q_2)}j_0^\dagger & 0 & 0 \\ -e^{-i(q_1+q_2)}j_0 & 3\mathbb{I} & 0 & 0 \\ 0 & 0 & 3\mathbb{I} & -e^{i(q_1+q_2)}j_0^\dagger \\ 0 & 0 & -e^{-i(q_1+q_2)}j_0 & 3\mathbb{I} \end{pmatrix}$$

and

$$M_{th} = \begin{pmatrix} j_0 u & j_0 v \\ -j_0^\dagger u & -j_0^\dagger v \\ j_0 w & 0 \\ -j_0^\dagger w & 0 \end{pmatrix}.$$

8

The entries are themselves matrices:

$$j_0 = \begin{pmatrix} 1 & 1 & 1 \\ e^{i(q_1+q_2)} & 1 & e^{iq_1} \\ e^{i(q_1+q_2)} & e^{iq_2} & 1 \end{pmatrix}$$

$$u = \cos\beta \begin{pmatrix} \sin\alpha & 0 & 0 \\ 0 & \sin(\alpha + \frac{4\pi}{3}) & 0 \\ 0 & 0 & \sin(\alpha + \frac{2\pi}{3}) \end{pmatrix}$$

$$v = -\sin\beta \begin{pmatrix} 1 & 0 & 0 \\ 0 & 1 & 0 \\ 0 & 0 & 1 \end{pmatrix}$$

$$w = \begin{pmatrix} \cos\alpha & 0 & 0 \\ 0 & \cos(\alpha + \frac{4\pi}{3}) & 0 \\ 0 & 0 & \cos(\alpha + \frac{2\pi}{3}) \end{pmatrix}.$$

The quantity $\psi$ describing the fluctuation amplitudes is given by

$$\psi(Q_1, Q_2) = \left(\rho_y, \sigma_y, \tau_y, \rho_z, \sigma_z, \tau_z, \mu_1^A, \mu_1^B, \mu_1^C, \nu_1^A, \nu_1^B, \nu_1^C, \mu_2^A, \mu_2^B, \mu_2^C, \nu_2^A, \nu_2^B, \nu_2^C\right).$$

Finally, the correction to the free energy due to the fluctuations in a given ground state, characterized by $\beta$ and $\alpha$, is computed by the formula

$$\delta F = T \int d^2q \log \Delta$$

with determinant

$$\Delta = \det \begin{pmatrix} J_{tt} M_t & J_{th} M_{th}^\dagger \\ J_{th} M_{th} & J_{hh} M_h \end{pmatrix}.$$

A straightforward analytical evaluation of this determinant is complicated by the large dimension of the matrices. However, we may still carry out this computation by making repeated use of the block structure of $\delta H$. We use the following block matrix decomposition:

$$\begin{pmatrix} A & B \\ C & D \end{pmatrix} = \begin{pmatrix} 1 & 0 \\ 0 & D \end{pmatrix} \begin{pmatrix} A & B \\ D^{-1}C & 1 \end{pmatrix} = \begin{pmatrix} 1 & 0 \\ 0 & D \end{pmatrix} \begin{pmatrix} 1 & B \\ 0 & 1 \end{pmatrix} \begin{pmatrix} A - BD^{-1}C & 0 \\ D^{-1}C & 1 \end{pmatrix}.$$

Since all three matrices are triangular, the determinant is just

$$\det \begin{pmatrix} A & B \\ C & D \end{pmatrix} = \det D \det\left(A - BD^{-1}C\right),$$

an equality known as the Schur identity. Iterative application of this identity to the block matrices breaks the block structure down, until we are left with a determinant of a $3 \times 3$ matrix. We start with the calculation of the determinant for the special case $\beta = \frac{\pi}{2}$. Here the determinant (after discarding $\alpha$-independent factors) can be written as

$$\sum_{n=0}^{3} \Delta_n \left(\frac{J_{th}}{\bar{J}}\right)^{2n}$$

with $\bar{J} = \sqrt{J_{tt} J_{hh}}$ and $q$-dependent $\Delta_n$. Computation of the $\Delta_n$ shows that only $n = 3$ yields an $\alpha$-dependence. The two other determinants are independent of $\alpha$ and are therefore not responsible for the selection of the groundstate. Integration over $q$ yields the sought free energy dependence on $\alpha$. Numerical evaluation of this last integral yields

$$\delta F \equiv \left(\frac{J_{th}}{\bar{J}}\right)^6 \frac{NT}{3} \int \frac{d^2q}{(2\pi)^2} \frac{\Delta_3}{\Delta_0} = -10^{-4} NT \cos^2(3\alpha) \left(\frac{J_{th}}{\bar{J}}\right)^6.$$

Next we compute to lowest-order in $\left(\frac{J_{th}}{\bar{J}}\right)$ the dependence on $\beta$. In computing the determinant by Schur's identity, one is left with a determinant of the form

$$\log \det \left[ \mathbb{I} + \left(\frac{J_{th}}{\bar{J}}\right)^2 A^{-1}B + \left(\frac{J_{th}}{\bar{J}}\right)^4 A^{-1}C \right],$$

where $A, B, C$ are $3 \times 3$ matrices. We are only interested in the lowest-order correction, which will obviously be proportional to $\left(\frac{J_{th}}{\bar{J}}\right)^2$. We find this most conveniently by rewriting the logarithm as a trace:

$$\log \det \left[ \mathbb{I} + \left(\frac{J_{th}}{\bar{J}}\right)^2 A^{-1}B + \left(\frac{J_{th}}{\bar{J}}\right)^4 A^{-1}C \right] = \text{Tr}[\log(\mathbb{I} + \left(\frac{J_{th}}{\bar{J}}\right)^2 A^{-1}B + \left(\frac{J_{th}}{\bar{J}}\right)^4 A^{-1}C)] \equiv \sum T_n \left(\frac{J_{th}}{\bar{J}}\right)^{2n}$$

Once the $T_n$ are known, the free energy can be expressed as

$$\delta F \equiv \sum_{n=1}^{6} \delta F_n \left(\frac{J_{th}}{\bar{J}}\right)^{2n} = \frac{NT}{3} \sum_{n=1}^{6} \left(\frac{J_{th}}{\bar{J}}\right)^{2n} \int \frac{d^2q}{(2\pi)^2} T_n$$

Next we compute the $T_n$ by expanding the logarithm in $\left(\frac{J_{th}}{\bar{J}}\right)^2$

$$\text{Tr}[\log(\mathbb{I} + \left(\frac{J_{th}}{\bar{J}}\right)^2 A^{-1}B + \left(\frac{J_{th}}{\bar{J}}\right)^4 A^{-1}C)] = \sum_{1}^{\infty} (-)^{n-1} \frac{\left(\frac{J_{th}}{\bar{J}}\right)^{2n} \text{Tr}\left[(A^{-1}B + \left(\frac{J_{th}}{\bar{J}}\right)^2 A^{-1}C)^n\right]}{n}$$

with

$$T_1 = \text{Tr}\left[A^{-1}B\right]$$
$$T_2 = \text{Tr}\left[A^{-1}C\right] - \frac{1}{2}\text{Tr}\left[(A^{-1}B)^2\right]$$
$$T_3 = -\text{Tr}\left[A^{-1}BA^{-1}C\right] + \frac{1}{3}\text{Tr}\left[(A^{-1}B)^3\right].$$

Here we will only require $T_1$. Using the stated formula, it is straightforward to calculate the lowest-order dependence of the free energy on $\beta$. We find

$$\delta F = 0.131274 \cos(2\beta) NT \left(\frac{J_{th}}{\bar{J}}\right)^2.$$

---

[1] J. Villain, J. Phys France **38**, 385 (1977).
[2] E. Shender, Sov. Phys. JETP **56**, 178 (1982).
[3] C. L. Henley, Phys. Rev. Lett. **62**, 2056 (1989).
[4] J. T. Chalker, P. C. W. Holdsworth, and E. F. Shender, Phys. Rev. Lett. **68**, 855 (1992).




\end{document}